\documentclass[prl,twocolumn,showpacs,amsmath,amssymb,superscriptaddress,unsortedaddress,floatfix]{revtex4}

\usepackage{pstricks}
\usepackage{subfigure}
\usepackage{graphicx}

\newcommand{\bra}[1]{\langle #1 |}
\newcommand{\ket}[1]{| #1 \rangle}

\newcommand{\pro}[1]{\ket{#1}\bra{#1}}
\newcommand{\unidim}[2]{\ket{#1}\bra{#2}}
\def\beq{\begin{equation}}
\def\eeq{\end{equation}}
\def\id{{\rm id}}

\begin{document}

\title{Entanglement-swapping boxes and their communication properties}

\author{Andrzej Grudka}
\affiliation{Institute of Theoretical Physics and Astrophysics,
University of Gda\'{n}sk, 80-952 Gda\'{n}sk, Poland}

\affiliation{National Quantum Information Centre of Gda\'{n}sk, 81-824 Sopot, Poland}

\affiliation{Faculty of Physics, Adam Mickiewicz University, 61-614
Pozna\'{n}, Poland}

\author{Micha{\l} Horodecki}

\affiliation{Institute of Theoretical Physics and Astrophysics,
University of Gda\'{n}sk, 80-952 Gda\'{n}sk, Poland}

\affiliation{National Quantum Information Centre of Gda\'{n}sk, 81-824 Sopot, Poland}

\author{Pawe{\l} Horodecki}

\affiliation{Faculty of Applied Physics and Mathematics, Technical
University of Gda\'{n}sk, 80-952 Gda\'{n}sk, Poland}

\affiliation{National Quantum Information Centre of Gda\'{n}sk, 81-824 Sopot, Poland}

\author{Ryszard Horodecki}

\affiliation{Institute of Theoretical Physics and Astrophysics,
University of Gda\'{n}sk, 80-952 Gda\'{n}sk, Poland}

\affiliation{National Quantum Information Centre of Gda\'{n}sk, 81-824 Sopot, Poland}

\author{Marco Piani}

\affiliation{Institute of Theoretical Physics and Astrophysics,
University of Gda\'{n}sk, 80-952 Gda\'{n}sk, Poland}

\affiliation{Faculty of Applied Physics and Mathematics, Technical
University of Gda\'{n}sk, 80-952 Gda\'{n}sk, Poland}

\date{\today}

\begin{abstract}
We  pose the fundamental question of communication properties of primitives irrespectively of their implementation. To illustrate the idea we introduce the concept of entanglement-swapping boxes, i.e. we consider any quantum operations which perform entanglement swapping,  not necessarily via simple quantum teleportation. We ask a question about the properties of such boxes., i.e. what local operations and how much classical communication  are needed to perform them. We also ask if any box which performs entanglement swapping can be used to establish classical communication. We show that each box needs at least  two bits of classical communication to perform it. It  is also shown that each box can be used for classical communication and, most importantly, that there exist boxes which allow to communicate at most  one bit. Surprisingly we find basic irreversibility in the process of entanglement swapping  with respect to its communication properties.
\end{abstract}


\maketitle

Classical communication properties of quantum bipartite operations have recently been studied (\cite{capacityBennett,twoway} and references therein). More precisely, the amount of classical communication (bits) that two parties can exchange by exploiting a given bipartite operation was estimated. The converse question, how much classical communication a given bipartite operation requires to be implemented, was also addressed~\cite{beckman}.  



However, as far as we know, only communication properties of {\emph specifically implemented}
bipartite quantum operations -- completely-positive trace-preserving maps -- have been  analyzed (e.g. one considers a particular protocol and asks the  questions:  \emph{How much communication is needed to perform it? How much communication classical or quantum can be established with it?}) 

In this article we ask another fundamental question:  \emph{What are
the communication properties of fundamental tasks, or primitives, irrespectively of their
implementation?} Indeed, it is often useful to formulate information processing in terms of primitives rather than via specific realizations (e.g. a particular protocol). 

In general, one can consider ``black boxes'' --  quantum operations whose details we may not know -- which realize a primitive, and analyze  the requirements of their inner implementation (the general structure of protocols, i.e. basic quantum operations which can be used such as unitaries, projectors and so on). We investigate for what  other tasks they can serve (e.g., if they can be used to establish classical communication).
We calculate how much classical communication one needs to implement a particular box  and how much classical communication one can establish with this box.  We stress that the box is designed to perform some precise task and the possibility of establishing classical communication is only a bonus.  We investigate if any box can give us such a bonus, e.g., if there are worst cases where classical communication is not possible at all.

As an example of general interest in the just described framework we introduce the concept of entanglement-swapping (ES) boxes associated to entanglement swapping~\cite{teleportation,swapping}. Entanglement swapping is a process where quantum correlations are distributed by means of (tripartite) Local Operations and Classical Communication (LOCC), and it is fundamental for, e.g., long distance quantum communication based on quantum repeaters~\cite{repeaters}.   We define the \emph{communication cost} (CC) and the \emph{communication value} (CV) of a \emph{particular} box,  i.e., one of the many possible boxes which realize a certain primitive, as the least amount of communication required to implement the box, i.e., the protocol, and allowed by the box itself, respectively~\footnote{We will refer to the number of bits of communication required (allowed) in the asymptotic limit of many instances of the box to be implemented (exploited).}. 
We further define the CC (CV) of a primitive as the minimal CC (CV) over all boxes realizing the primitive, i.e. as the minimal communication that we have to spend to realize (that we can get from) \emph{any} black box that we know to perform the primitive. Surprisingly, we show that  communication has to be spent to perform a primitive that can be established with it. Hence the process of ES is in general irreversible  with respect to classical communication.  A similar irreversibility occurs for Popescu-Rohrlich boxes \cite{PR-machine}, where one needs one bit of classical communication to establish correlations, however this bit cannot be used to send a message, i.e. Popescu-Rohrlich boxes are nonsignaling \cite{barrett2005, toner-bacon, degorrePRA2005, CerfGMP2005, beckman, nsb-phhh}.

\begin{figure}[!t]
\includegraphics[width=0.35\textwidth]{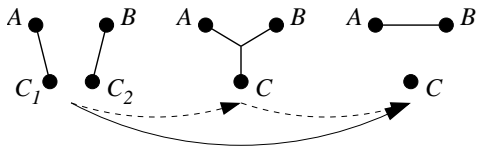}
\caption{Entanglement swapping (continuous arrow) in two steps (dashed arrows): from two EPR pairs to a GHZ state, to the final EPR pair. Lines denote quantum correlations.}
\label{fig:swappingsteps}
\end{figure}

We also report results about boxes associated to other primitives, namely two intermediate steps of ES: creating a GHZ state from two Bell states  and creating a Bell state between any two parties starting from a GHZ 
state (Figure \ref{fig:swappingsteps}). 

\begin{figure}
\subfigure[~Defining action of an ES-box. Time goes from left to right.]{\label{fig:ESbb}\includegraphics{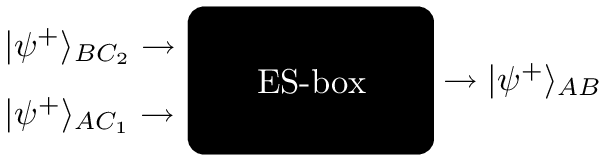}}
\subfigure[~Actual (LOCC) implementation of a given ES-box.]{\label{fig:ESb-inside}\includegraphics{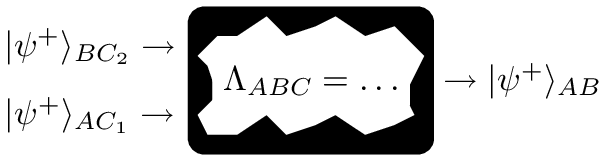}}
\caption{Entanglement swapping (black) box.} 
\end{figure}

Imagine that Alice and Bob share maximally entangled states $|\Psi^{+}\rangle_{AC_{1}}$ and $|\Psi^{+}\rangle_{AC_{2}}$ with (two subsystems of) Charlie. ES corresponds to obtaining, from such initial resource, a state $|\Psi^{+}\rangle_{AB}$, by LOCC. We define a (black) ES-box to be any tripartite-LOCC operation
such that ES takes place upon the input of $|\Psi^{+}\rangle_{AC_{1}}|\Psi^{+}\rangle_{BC_{2}}$ (see Figure \ref{fig:ESbb}). We will typically refer to the $AB$ output alone, disregarding the output subsystems of Charlie, i.e. effectively tracing them out.

Because of the characterization of all ES-boxes we will provide, the CC and the CV will be analyzed with respect to the relevant direction $C\rightarrow AB$. If not differently stated, we will allow Alice and Bob to act together when considering the CV. Indeed, while the box is assumed to be (in its \emph{implementation}) tripartite LOCC, this does not force the \emph{users} to have the same limitation. Alternatively, one could think of entanglement between Alice and Bob being consumed while trying to get a signal from Charlie by means of the box.

We may imagine that a provider, Paul, sells ES-boxes -- a tool to process some precious resource like pre-shared entanglement -- to the users Alice, Bob and Charlie.
Suppose now that Paul wants to minimize the use of the boxes for any other purpose than just
entanglement swapping. In particular, he might like to prevent the
parties from using the box to communicate. Thus, he tries to
minimize the CV of the box. On the other
hand, he would like to built the box in the cheapest way, hence
he tries to minimize the CC of the box. Among all ES-boxes, we will be interested in the best possible boxes
from the point of view of the provider, i.e. the ones with the minimal CC and CV, therefore in the CC and CV of the primitive as defined above.

We start by noticing that any ES box $\Lambda_{ABC}$ is {\it signaling}
with respect to $C \rightarrow AB$~\cite{beckman,nsb-phhh}. In order to
see this, we have to show that there exist some initial state
$\varrho_{ABC}$ and (at least two) operations on Charlie's subsystem
$\{\Gamma^{i}_{C}\}$ such that the states
$\varrho^{i}_{AB}=\Lambda_{ABC}[\Gamma^{i}_{C}(\varrho_{ABC})]$ are
different. To communicate by means of the state
$\varrho_{ABC}$ and of the box $\Lambda_{ABC}$, (i) Charlie applies one operation $\Gamma^{i}_{C}$
depending on the ``letter'' he wants to send, (ii) the parties let the box act,
and finally (iii) Alice and Bob try to guess which state
$\varrho^{i}_{AB}$ they have.  If the states $\varrho^{i}_{AB}$ are different, the
box has nonzero CV.  Let us choose the
initial state
$|\Psi^{+}\rangle_{AC_{1}}|\Psi^{+}\rangle_{BC_{2}}$,
and as Charlie's possible operations the identity and the measurement
of his two qubits performed in the computational basis. If Charlie applies
the identity -- does nothing -- then, by definition of ES-box, Alice and Bob obtain as
output the state $|\Psi^{+}_{AB}\rangle$.
On the other hand, the measurement by Charlie
destroys  all the entanglement between the parties, that can not be restored by the LOCC ES-box, thus Alice and Bob can at most obtain a separable state.

We now proceed with a specific example of ES-box. As our tripartite
LOCC map we take \emph{any} LOCC teleportation protocol~\cite{teleportation} from Charlie to Alice, i.e. such that an input state $\ket{\Psi^+}_{AC_1}\ket{\psi}_B\ket{\phi}_{C_2}$ is transformed into $\ket{\phi}_{A}\ket{\psi}_B$, for all $\ket{\phi}$. This particular ES-box has CV equal to $2$  via dense coding~\cite{BennettWiesner}, even considering just one final receiver, i.e. $C\rightarrow A$. Indeed, one can consider an initial state $\ket{\Psi^+}_{AC_1}\ket{\psi}_B\ket{\Psi^+}_{A'C_2}$ involving a local ancilla $A'$ of Alice that is maximally entangled with $C_2$. Charlie can locally rotate the pair $\ket{\Psi^+}_{A'C_2}$ into one of four orthogonal Bell states, which is fully in the hands of Alice after teleportation.  If we choose the standard teleportation protocol, a Bell measurement at the sender followed by a proper unitary rotation at the receiver, the CC is 2, and it could not be anyway less because of causality~\cite{teleportation}.

As we will show, there are ES boxes which have CV
equal to $1$. Then the simple argument of causality cannot be used
to prove that their CC is greater or equal to $2$. Anyway, we will show that every
ES-box has CC greater or equal to $2$. Thus, there are ES-boxes which exhibit
communication irreversibility: the box needs
more bits to be implemented than it can signal.

In the theorem below we characterize all LOCC maps that perform
entanglement swapping, i.e. every ``internal implementation'' of ES-boxes (Figure~\ref{fig:ESb-inside}).

{\bf Theorem 1}. \emph{Any ES-box $\Lambda_{ABC}$ is of the form
$\Lambda(\varrho_{ABC})={\rm Tr}_{C}\big(\sum_{i}U_{A}^{i}U_{B}^{i}E_{C}^{i}(\varrho_{ABC})U_{A}^{i\dagger}U_{B}^{i\dagger}E_{C}^{i\dagger}\big)$,
where $U_{A}^{i}$ and $U_{B}^{i}$ are unitary operations and
$E_{C}^{i}=\unidim{u^i}{\psi^i_+}$ are rank-one measurement operators, with $\ket{\psi^i_+}$ normalized maximally entangled states of Charlie's
particles, which satisfy
$\sum_{i}E_{C}^{i\dagger}E_{C}^{i}=\sum_i\|u_i\|^2\pro{\psi^i_+}=\openone_{C_1C_2}$.}

\emph{Proof.}  We provide here only a sketch of the full proof~\cite{longswapping}, which consists of two parts. In the first part \emph{(I)} we prove that Alice and Bob cannot perform non-unitary
operations. In the second part \emph{(II)} we find the conditions
which have to be satisfied by Charlie's operators.

\emph{(I)} We will prove this part by contradiction. Every LOCC protocol corresponds to rounds of local actions coordinated by classical communication, by means of which the results of the local operations are transmitted to the other parties. Without loss of generality, let Alice be the first, between her and Bob, to perform an operation different from an isometry.  The initial entanglement in the $A|BC$ splitting corresponds to one ebit, and the same holds at the end of the protocol by the ES condition. As the protocol is LOCC,  there must be one ebit of $A|BC$ entanglement at every step. One can prove that any operation by Alice that were not an isometry would violate such a condition. 
Therefore, neither Alice nor Bob can perform any other operation but isometries. Since their output subsystems correspond to their input subsystems, such isometries are indeed unitaries.

\emph{(II)} Charlie is initially maximally entangled
with Alice and Bob. At the end, Alice and Bob, who both can perform only
unitary operations, have a pure state. Therefore, Charlie has to disentangle his
qubits from Alice and Bob's, i.e. he has to perform a measurement with measurement
operators of rank $1$: $E_C^i=\unidim{u^i}{\psi^i}$, with $\ket{\psi_i}$ a normalized state. The output state of Alice's and Bob's subsystem corresponding to a given $E_C^i$ has the same Schmidt
coefficients as $\ket{\psi_i}$. Thus, we see that Alice and Bob may obtain a maximally
entangled state only if $\ket{\psi_i}$ is maximally entangled.\hfill$\square$

We can now address the problem of the CC of ES-boxes. Indeed, given the standard form for ES-boxes of Theorem 1, it is clear that all ES-boxes can be realized with  only $C\rightarrow AB$ classical communication, which is used by Charlie to tell Alice and Bob the result of his measurement, so that they can apply the correct local unitary rotations. We will need the entropic quantities $S(A|B)=S(AB)-S(B)$ (conditional entropy), $I(A:B)=S(A)+S(B)-S(AB)$ (mutual information), $I(A:B|R)=S(A|R)+S(B|R)-S(AB|R)$ (conditional mutual information), where for brevity we use the notation $S(X)=S(\varrho_X)$, etc., and the following lemma.

{\bf Lemma 1}. \emph{Consider an ensemble $\{p^{i},\varrho_{AB}^{i}\}$, and the corresponding average state $\varrho_{AB}=\sum_ip^i\varrho^i_{AB}$.
Then $\Delta I \leq \Delta S$, where $\Delta I= \sum_{i}p^{i} I
(\varrho_{AB}^{i})-I(\varrho_{AB})$ is the average increase of mutual
information, and $\Delta
S=S(\varrho_{AB})-\sum_{i}p^{i}S(\varrho_{AB}^{i})$ is the average decrease of
entropy, when Alice and Bob come to know the index of the state they actually share among the ones in the ensemble.} 

\emph{Proof}. Let us introduce an ancilllary system $R$ and consider the tripartite state $\varrho_{ABR}=\sum_{i}
p^{i} \varrho_{AB}^{i}\otimes|i \rangle_{R}\langle i|_{R}$, with orthogonal $\ket{i}_R$ on $R$. Note that
$\Delta I = I(A:B|R)-I(A:B)$ and $\Delta S = I(AB:R)$, with the right-hand sides of these equalities calculated with respect to $\varrho_{ABR}$. We have therefore
\begin{equation}
\Delta I - \Delta S=-I(A:R)-I(B:R)\leq 0,
\end{equation}
 because of positivity of mutual information. \hfill$\square$

The number of bits sent by Charlie to Alice and Bob per each realization of a ES-box can not be less than the Shannon entropy $H(\{p_i\})=-\sum_ip_i\log p_i$ of the probability distribution of the outcomes of Charlie.

{\bf Theorem 2}. \emph{Any ES-box has CC at least equal to 2.}

\emph{Proof.} After Charlie's measurement, Alice and Bob have
an ensemble $\{p^{i},\varrho_{AB}^{i}\}$. Each $\varrho_{AB}^i$ is a (different) maximally entangled state, and $\varrho_{AB}=\sum_ip^i\varrho_{AB}^i=\openone_4/4$ corresponds to Alice and Bob's initial reduced state. From Lemma 1 we have
\beq
\begin{split}
H(\{p_{i}\})&\geq S(\varrho_{AB})-\sum_{i}p^{i}S(\varrho_{AB}^{i})\\
&\geq\sum_{i}p^{i}I(\varrho_{AB}^{i})-I(\varrho_{AB})=2.
\end{split}
\eeq
\hfill$\square$

Since we know that for the standard teleportation map the CC is equal to $2$, by using
Theorem 2 we obtain that CC of ES is equal to $2$.

Let us now consider the CV of a ES box: the theorem below
gives a lower bound valid for all ES-boxes.

{\bf Theorem 3}. \emph{Any ES-box has CV at least equal to 1.}

\emph{Proof}. One can check that if the initial state $\varrho_{ABC}$ is
$|\Psi^{+}\rangle_{AC_{1}}|\Psi^{+}\rangle_{BC_{2}}$,
and Charlie's possible operations $\{\Gamma^{i}\}$ before the ES-box correspond to the set of unitaries $S=\{\openone_{C_1} \otimes \openone_{C_2},Z_{C_1}
\otimes \openone_{C_2}\}$, then the states
$\{\varrho^{i}_{AB}=\Lambda_{ABC}[\Gamma^{i}_{C}(\varrho_{ABC})]\}$
are orthogonal and can be perfectly distinguished by Alice and Bob.
\hfill$\square$
It should be emphasized that, in order to communicate, Charlie can apply operations $\{\Gamma^{i}\}$ that do not depend on the ES-box, so that communication is achieved whatever the \emph{black} ES-box (see Figure~\ref{fig:ESbb}) at disposal, i.e. the internal structure of the
box -- the particular LOCC map $\Lambda_{ABC}$ -- is not relevant.
We now prove that the bound CV=1 can be achieved. The maps we provide to this purpose happen to be also $C \rightarrow A$ and $C
\rightarrow B$ {\it nonsignaling}, i.e. Charlie cannot communicate 
to neither Alice nor Bob separately. We will use the standard bipartite operation of $UU^*$-twirling $\Lambda_{AB}^{T}(\sigma_{AB})\equiv\int dU(U_A\otimes U^{\ast}_B)\sigma_{AB} (U_A \otimes U^{\ast}_B)^{\dag}$, where $\int dU$ denotes integration over the unitary group with respect to the Haar measure, and $U^\ast$ is the complex conjugate of $U$.

{\bf Theorem 4}. \emph{Apply the $UU^*$-twirling to the output of any ES-box $\Lambda_{ABC}$: the resulting map is again an ES-box, with CV equal to $1$, and non-signaling with respect to $C \rightarrow A$ and $C \rightarrow B$.}

\emph{Proof.} Consider the map $\tilde{\Lambda}_{ABC}=\Lambda_{AB}^{T}\circ\Lambda_{ABC}$. $\tilde{\Lambda}_{ABC}$ is obviously a ES-box, since $\Lambda_{AB}^{T}$ leaves $\ket{\Psi^+}$ invariant. To provide an upper bound, that takes care of the possibility of exploiting pre-established quantum correlations, on the CV of $\tilde{\Lambda}_{ABC}$, we use the entanglement assisted classical capacity of a quantum channel (EACCQC)~\cite{EACapacity}. Consistently with the idea of obtaining an upper bound, we apply the formula for EACCQC by pretending that Charlie has complete control over all the input -- not only on its subsystem, so that it reads
\beq
\begin{split}
C(\tilde\Lambda_{ABC})=\max_{\varrho_{ABC}} \Big(&S(\varrho_{ABC}) +
S\big(\tilde\Lambda_{ABC}(\varrho_{ABC})\big)\\-&S\big((\tilde\Lambda_{ABC}\otimes \id_{E}\big)(\Phi_{ABCE})\big)\Big)
\end{split}
\eeq
where $\Phi_{ABCE}$ is any purification of $\varrho_{ABC}$, and the maximum
is taken over all input states $\varrho_{ABC}$. For our channel we have
$\tilde{\Lambda}_{ABC}(\varrho_{ABC})=F|\Psi^{+}\rangle_{AB} \langle
\Psi^{+}|_{AB} +\frac{1-F}{3}(\openone_{AB}-|\Psi^{+}\rangle_{AB} \langle
\Psi^{+}|_{AB})$ and $(\tilde{\Lambda}_{ABC}\otimes
\id_{E})(\Phi_{ABCE})=F|\Psi^{+}\rangle_{AB} \langle \Psi^{+}|_{AB}
\otimes \varrho^{0}_{E} +\frac{1-F}{3}(\openone_{AB}-|\Psi^{+}\rangle_{AB}
\langle \Psi^{+}|_{AB}) \otimes \varrho^{1}_{E}$, with $0\leq F\leq1$. 
 After some algebra, we obtain
\[
\begin{split}
C(\tilde{\Lambda}_{ABC})= \max
\Big(&S\big(F\varrho^{0}_{E}+(1-F)\varrho^{1}_{E}\big)\\
- &\big(FS(\varrho^{0}_{E})+(1-F) S(\varrho^{1}_{E})\big)\Big) \leq 1.
\end{split}
\]
On the other hand, from Theorem 3 we know that any ES box has
CV at least 1, therefore we find that the CV of $\tilde{\Lambda}_{ABC}$ is 1.
Moreover, $\text{Tr}_{B(A)}(\tilde\Lambda_{ABC}(\varrho_{ABC}))=\frac{1}{2}\openone_{A(B)}$, for whatever $\varrho_{ABC}$, therefore $\tilde\Lambda_{ABC}$ is $C \rightarrow A(B)$ {\it nonsignaling}. \hfill$\square$

From Theorems 3 and 4 we conclude that ES has CV equal to 1. We note also that a map $\tilde\Lambda_{ABC}$, as defined in Theorem 4, has CC less or equal to that of the map $\Lambda_{ABC}$ from which it derives. This follows from the fact that twirling can be performed  without communication by means of shared randomness. We may take $\Lambda_{ABC}$ to be standard teleportation, so that $\tilde\Lambda_{ABC}$ has CC equal to 2 and CV equal to 1. Thus both CC and CV of this specific ES-box coincide with those of ES.

One may say that ES is an irreversible primitive from the point of view of classical communication, because the minimal amount of communication required by any ES-box is higher than the minimal amount of communication that we can expect to obtain from a ES-box. Indeed, there are specific ES-boxes which incarnate this irreversibility.

 As a primitive, ES may be split into two subprimitives, or intermediate steps (Figure \ref{fig:swappingsteps}): the transformation of two EPR pairs into a GHZ state, and of the latter into an EPR pair between any two subsystems.  We remark that this does not mean that every ES-box can be split correspondingly into two sub-boxes. By essentially the same methods used for the full ES, one can prove~\cite{longswapping} that both the CC and CV of the second (sub)primitive are equal to $1$, so that such primitive is \emph{reversible} from the point of view of classical communication. With respect to the first subprimitive, one can  show~\cite{longswapping} that there exists a  corresponding box which has CC equal to $1$ and CV strictly less than $1$. Moreover this submprimitive has CV strictly positive. In order to prove this last statement, let us notice that after the action of the box on $|\Psi^{+}\rangle_{AC_{1}}|\Psi^{+}\rangle_{BC_{2}}$, Alice and Bob's reduced density matrix is $\varrho_{AB}=\frac{1}{2}(|00\rangle \langle00| + |11\rangle \langle11|)$. We now argue that, before the box is applied, Charlie can transform the input state in such a way that the output of the box yields a different reduced density matrix for Alice and Bob. To this end, Charlie applies a random unitary operations which produce the state $\varrho_{ABC}=\frac{1}{4}I_{AB} \otimes \frac{1}{4}I_{C}$.  As the (conditional) action of the box on  Alice's and Bob's systems consists of unitary operations, it cannot change Alice's and Bob's reduced density matrix $\varrho_{AB}=\frac{1}{4}I_{AB}$, which commutes with all unitaries. Since CV for the first subprimitive is strictly greater than 0 and CV for the second subprimitive as well as for the  ``composed'' ES-box is equal to 1,  we conclude that CV for primitives is strictly subadditive. On the other hand CC is additive.

In this article we have considered communication properties of primitives irrespectively of their implementation. We exemplified this by introducing the concept of entanglement-swapping boxes, and characterizing them with respect to LOCC operations.  We studied their classical-communication properties, and discovered a phenomenon of irreversibility of the primitive itself: it needs more bits to be implemented than it can signal. We also reported results of a similar analysis for primitives corresponding to intermediate steps of entanglement swapping. We believe that studying communication properties of primitives will lead to a better understanding of the interplay between the quantum and the classical,  at the level both of properties and of interconversion of resources.

We thank K. Horodecki for useful discussions.
This work was supported by the European Commission through the Integrated Project FET/QIPC ``SCALA''. AG was also supported by grant 1 P03B 014 30 of the State Committee for
Scientific Research.

\end{document}